\documentclass[pre,twocolumn,showpacs,showkeys,superscriptaddress,preprintnumbers,floatfix]{revtex4-1}

\usepackage{etex}
\usepackage{ifpdf}
\usepackage{hyperref}
\usepackage{dcolumn}
\usepackage{url}
\usepackage{amsmath}
\usepackage{amscd}
\usepackage{amsfonts}
\usepackage{amssymb}
\usepackage{amsthm}
\usepackage{xfrac}
\usepackage{braket}
\usepackage{bm}   % bold math
\usepackage{bbm}
\usepackage{verbatim}
\usepackage{mathtools}
\usepackage{stmaryrd}
\usepackage{xcolor}
\usepackage{setspace}
\usepackage{tikz}
\usetikzlibrary{arrows}
\usetikzlibrary{automata}
\usetikzlibrary{calc}
\usetikzlibrary{decorations.pathreplacing}
\usetikzlibrary{patterns}
\usetikzlibrary{positioning}
\usetikzlibrary{plotmarks}
\usetikzlibrary{shapes}

\usepackage{pgfplots}
\pgfplotsset{compat=newest}

\tikzstyle{vaucanson}=[
  node distance=2.5cm,
  bend angle=15,
  auto,
  every loop/.style={},
  every edge/.style={->,draw=black,line width=1.2,>=latex,shorten <=1pt,shorten >=1pt},
  every state/.style={draw=black,line width=2,font=\large},
  loop right/.style={right,out=22,in=-22,loop},
  loop above/.style={above,out=112,in=68,loop},
  loop left/.style={left,out=202,in=158,loop},
  loop below/.style={below,out=292,in=248,loop},
  loop above right/.style={right,out=67,in=23,loop},
  loop above left/.style={left,out=157,in=113,loop},
  loop below left/.style={left,out=247,in=203,loop},
  loop below right/.style={right,out=337,in=293,loop},
]

\theoremstyle{plain}    
\theoremstyle{plain}    
\theoremstyle{plain}    
\theoremstyle{plain}    
\theoremstyle{plain}    
\theoremstyle{plain}    
\theoremstyle{plain}    
\theoremstyle{plain}    
\theoremstyle{plain}    
\theoremstyle{plain}    
\theoremstyle{plain}    
\theoremstyle{plain}    
\theoremstyle{plain}    
\theoremstyle{plain}    
\theoremstyle{plain}    
\theoremstyle{plain}    
\theoremstyle{plain}

\newcommand{\MeasAlphabet}  {\mathcal{A}}
\newcommand{\MeasSymbol}   { {X} }
\newcommand{\meassymbol}   { {x} }

\newcommand{\CausalState}   { \mathcal{S} }

\newcommand{\causalstate}   { \sigma }
\newcommand{\CausalStateSet}    { \bm{\CausalState} }
\newcommand{\AlternateState}    { \mathcal{R} }

\newcommand{\AlternateStateSet} { \bm{\AlternateState} }

\newcommand{\EE}        {{\bf E}}

\newcommand{\ProcessAlphabet}   {\MeasAlphabet}

\newcommand{\forward}{+}
\newcommand{\reverse}{-}
\newcommand{\forwardreverse}{\pm} % \pm

\newcommand{\FutureCausalState} { {\CausalState}^{\forward} }

\newcommand{\PastCausalState}   { {\CausalState}^{\reverse} }

\newcommand{\lastindex}[2]{
  \edef\tempa{0}
  \edef\tempb{#2}
  \ifx\tempa\tempb
    \edef\tempc{#1}
  \else
    \edef\tempa{0}
    \edef\tempb{#1}
    \ifx\tempa\tempb
      \edef\tempc{#2}
    \else
      \edef\tempc{#1+#2}
    \fi
  \fi
  \tempc
}

\newcommand{\CSjoint}[1][,]{
   \edef\tempa{:}
   \edef\tempb{#1}
   \ifx\tempa\tempb
      \ensuremath{\FutureCausalState\!#1\PastCausalState}
   \else
      \ensuremath{\FutureCausalState#1\PastCausalState}
   \fi
}

\newif\ifpm
\edef\tempa{\forwardreverse}
\edef\tempb{\pm}
\ifx\tempa\tempb
   \pmfalse
\else
   \pmtrue
\fi

\colorlet {R_color}    {blue}
\colorlet {k_color}    {black!30!green}

\def\clap#1{\hbox to 0pt{\hss#1\hss}}

\begin{document}

\title{Comment on Deterministic Information Bottleneck}

\author{Sarah E. Marzen}
\email{smarzen@cmc.edu}
\affiliation{W. M. Keck Science Department, Claremont CA 91711}

\date{\today}
\bibliographystyle{unsrt}

% ************************* ABSTRACT *************************
\begin{abstract}
We make the case that although Deterministic Information Bottleneck may be a contribution to clustering, it should not be used to aid lossy compression without the addition of blocklength. We therefore suggest a new objective function that does so and leave its testing to future work.
\end{abstract}


\preprint{arxiv.org:1702.08565 [physics.gen-ph]}

\maketitle

% ****************************************************************

%\tableofcontents
\setstretch{1.1}

% Handy abbreviations in the following
\newcommand{\Abet}{\ProcessAlphabet}
\newcommand{\MS}{\MeasSymbol}
\newcommand{\ms}{\meassymbol}
\newcommand{\SSet}{\CausalStateSet}
\newcommand{\St}{\CausalState}
\newcommand{\st}{\causalstate}
\newcommand{\MxSt}{\AlternateState}
\newcommand{\MxSSet}{\AlternateStateSet}
\newcommand{\mxst}{\mu}
\newcommand{\mxstt}[1]{\mu_{#1}}
\newcommand{\StartMS}{\bra{\delta_\pi}}
\newcommand{\Ipred}{\EE}
\newcommand{\ISI} { \xi }

\newcommand{\ECT}{\widehat{\EE}}
\newcommand{\CCT}{\widehat{C}_\mu}

\newcommand{\gen}{g}
\newcommand{\FeatAlphabet}{\mathcal{F}}

%%%%%%%%%%%%%%%%%%%%%%%%%%%%%%%%%%%%%%%%%%

\vspace{0.2in}

In $2017$, the ``Deterministic Information Bottleneck'' (DIB) method was published and is now highly cited. The basic idea is that according to the information bottleneck method \cite{tishby2000information}, an extension of rate-distortion theory \cite{berger} to an informational distortion measure, one should minimize expected distortion and also minimize the mutual information between the compressed signal and the original variable at the same time. The information bottleneck method is great for clustering, e.g. as in Ref. \cite{still2003geometric}, and it also is a contribution to information theory \cite{gastpar2003code}, but if one is not careful, one might misunderstand where the objective function comes from. In this short arXiv paper, we simply aim to describe where these rate-distortion objectives come from and why DIB is a contribution to clustering but not to information theory as it stands. We then modify the objective so that it might be a contribution to information theory and leave the testing for later work.

From one of my own papers \cite{marzen2024resource}: ``In the classic rate-distortion setup, one sends a sequence of $n$ letters $x_{0:n}$ to an encoder that chooses one of $M$ words for those $n$ letters and then sends that word to a decoder which produces a guess as to what those letters were, $\hat{x}_{0:n}$. The material constraint is actually $\log M/n$, not a mutual information. This corresponds to a more intuitive notion of resource constraints in the biological sense-- number of molecules or number of neurons, normalized by ``blocklength'' $n$. Some distortion measure is defined, $d(x,\hat{x})$, which in some extensions can be a distortion of the entire block $x_{0:n}$ relative to $\hat{x}_{0:n}$ rather than letter-by-letter. There are some rates $\log M/n$ and distortions $\sum d(x_i,\hat{x}_i)/n$ that are achievable and some that are unachievable given any combination of encoder and decoder. A theorem shows that the curve separating achievable from unachievable is given by replacing the rate $\log M/n$ with a mutual information $I[X;\hat{X}]$ and the average distortion with an expected distortion if all is memoryless. This curve is accurate in the limit that blocklength $n$ goes to infinity. Otherwise, the rate-distortion curve that separates achievable from unachievable is given by $R_n(D)$ rather than $R(D)$, and $R_n(D)$ is horribly difficult to calculate \cite{berger}.''

In the information bottleneck method, one replaces $d(x,\hat{x})$ with an informational distortion measure (a Kullback-Leibler divergence), which does not correspond to the classic rate-distortion setup, in which $d(\hat{x},x)$ does not depend on $P(Y|\hat{X}=\hat{x})$. As a result, the objective in the information bottleneck method becomes $I[R;Y]-\beta I[X;R]$, where $R$ is the compressed variable, $X$ is the variable that we compress, and $Y$ is the variable that we try to retain information about. There is a rate-distortion theorem justifying this objective \cite{gastpar2003code}.

The wonderful thing about the rate-distortion theorem is that it allows one to test if a sensor is near-optimal. One calculates the rate and expected distortion and places it relative to the rate-distortion curve, as in Ref. \cite{palmer2015predictive}. It is also known that the actual soft clusters obtained in order to get the rate-distortion curve are terrible for coding, as their coding rate is given by $H[R]$ rather than by $I[R;X]$ \cite{berger}, although they may be useful for machine learning as stated earlier. As such, they should not be used as a null model to decide if an information-theoretic objective is successful at producing good codes.

DIB seems to be a fix to the need for clusters with a good coding rate. In DIB, we replace $I[X;R]$ with $H[R]+\lambda H[R|X]$ for some other $\lambda$ and we take the limit of determinism, where $\lambda$ becomes very large. However, there is no rate-distortion theorem justifying this objective as revealing sensors that are closer to optimal, and in fact, there should not be, as the objective that reveals if a sensor is close to the boundary between achievable and unachievable is the original information bottleneck objective function.

We might ask-- does DIB lead to good information theory codes? The objective indeed leads to deterministic codes. Deterministic codes are what one would desire for lossy compression, but it is well-known that the best codes often require large blocklengths $n$, though there are exceptions. We therefore simply propose that perhaps the objective function of $\langle d(\hat{x}_{0:n},x_{0:n})\rangle+\beta H[R]+ \alpha H[R|X_{0:n}]-\lambda H[\hat{X}_{0:n}|R]$ should be investigated for information theorists, despite the computational overhead that is incurred by including blocklength.

All of this does not mean that DIB is not great at getting interesting clusters for machine learning and data science purposes.

%=================================================================
% References:
%=================================================================
% Use the following option to include external BibTeX files:
\bibliography{chaos}

\begin{thebibliography}{1}

\bibitem{tishby2000information}
Naftali Tishby, Fernando~C Pereira, and William Bialek.
\newblock The information bottleneck method.
\newblock {\em arXiv preprint physics/0004057}, 2000.

\bibitem{berger}
Thomas Berger.
\newblock {\em Rate distortion theory: A mathematical basis for data
  compression}.
\newblock Prentice-Hall, Inc., 1971.

\bibitem{still2003geometric}
Susanne Still, William Bialek, and L{\'e}on Bottou.
\newblock Geometric clustering using the information bottleneck method.
\newblock {\em Advances in neural information processing systems}, 16, 2003.

\bibitem{gastpar2003code}
Michael Gastpar, Bixio Rimoldi, and Martin Vetterli.
\newblock To code, or not to code: Lossy source-channel communication
  revisited.
\newblock {\em IEEE Transactions on Information Theory}, 49(5):1147--1158,
  2003.

\bibitem{marzen2024resource}
Sarah Marzen.
\newblock Resource-rational reinforcement learning and sensorimotor causal
  states.
\newblock {\em arXiv preprint arXiv:2404.18775}, 2024.

\bibitem{palmer2015predictive}
Stephanie~E Palmer, Olivier Marre, Michael~J Berry, and William Bialek.
\newblock Predictive information in a sensory population.
\newblock {\em Proceedings of the National Academy of Sciences},
  112(22):6908--6913, 2015.

\end{thebibliography}

%\onecolumngrid
%\appendix
%\clearpage
%\input{supp_final}
%\clearpage

\end{document}